\documentstyle[12pt,epsf]{article}
\def\Journal#1#2#3#4{{#1}{\bf #2} (#3) #4}

\def\NIMA{{\rm Nucl. Instr. and Meth.} A}

\newcommand{\bc}{\begin{center}}
\newcommand{\ec}{\end{center}}
\newcommand{\bi}{\begin{itemize}}
\newcommand{\ei}{\end{itemize}}
\newcommand{\beq}{\begin{equation}}
\newcommand{\eeq}{\end{equation}}

\newcommand{\Zzero}{\mbox{$\mathrm{ Z}^0$}}

\newcommand{\ts}{\thinspace}
\newcommand{\etal}{{\it et al.}}

\newcommand{\twog} {\thinspace 2$\gamma$}

\newcommand{\gevocc} {\mbox{${\mathrm \thinspace GeV}/c^2$}}

\newcommand{\sm} {Standard Model}
\newcommand{\A} {\mbox{$\mathrm{A^0}$}}
\newcommand{\h} {\mbox{$\mathrm{h^0}$}}
\newcommand{\Hpm} {\mbox{$\mathrm{H^{\pm}}$}}
\newcommand{\Ho} {\mbox{$\mathrm{H^{0}}$}}
\newcommand{\ma} {\mbox{$m_{\mathrm A}$}}
\newcommand{\mh} {\mbox{$m_{\mathrm h}$}}
\newcommand{\mH} {\mbox{$m_{\mathrm H}$}}
\newcommand{\ecm}{\mbox{$E_{\mathrm CM}$}}
\newcommand{\tanb} {\mbox{$\tan\beta$}}
\newcommand{\ee}{\mbox{$\mathrm {e^+e^-}$}}
\newcommand{\mm}{\mbox{$\mathrm {\mu^+\mu^-}$}}
\newcommand{\llbar}{\mbox{$\mathrm {\ell^+ \ell^-}$}}
\newcommand{\nnbar}{\mbox{$\mathrm {\nu \bar \nu}$}}

\newcommand{\WW}{\mbox{$\mathrm {W^+W^-}$}}

\newcommand{\qqbar}{\mbox{$\mathrm {q \bar q}$}}
\newcommand{\bbbar}{\mbox{$\mathrm {b \bar b}$}}
\newcommand{\ccbar}{\mbox{$\mathrm {c \bar c}$}}
\newcommand{\ggbar}{\mbox{$\mathrm {gg}$}}

\newcommand{\ttbar}{\mbox{$\tau^+ \tau^-$}}
\newcommand{\pb}{\mbox{$\mathrm {pb}^{-1}$}}
\begin{document}
%%%%%%%%%%%%%%%%%%%%%%%%%%%%%%%%%%%%%%%%%%%%%%%%%%
\pagenumbering{roman}
\begin{titlepage}
\begin{center}{\large   EUROPEAN ORGANIZATION FOR NUCLEAR RESEARCH
}\end{center}\bigskip
\begin{flushright}
  CERN-EP/2002-058   \\
  19 July 2002
 \end{flushright}
\bigskip\bigskip
\begin{center}
{\huge\bf\boldmath Search for a low mass CP-odd Higgs boson
in $\mathrm{e^+e^-}$ collisions with the\unboldmath}\\
\vspace{0.3cm}
{\huge\bf\boldmath OPAL detector at LEP2 \unboldmath}
\end{center}\bigskip

\begin{center}
      {\Large\bf The Opal Collaboration }\\

\end{center}\bigskip

\bc
{\large  \bf Abstract}\\
\ec

We have analysed the data collected by OPAL at centre-of-mass
energies between 189 and 209 GeV searching for Higgs boson
candidates from the process $\ee \to \h\Zzero$ followed by the
decay of $\h \to \A\A$ where \A\ is the CP-odd Higgs boson. The
search is done in the region where the \A\ mass, \ma, is below the
production threshold for \bbbar, and the CP-even Higgs boson mass
\mh~is within the range 45--86 \gevocc. In this kinematic range,
the decay of $\h\to \A\A$ may be dominant and previous Higgs boson
searches have very small sensitivities. This search can be
interpreted within any model that predicts the existence of at
least one scalar and one pseudoscalar Higgs boson. No excess of
events is observed above the expected \sm\ backgrounds.
Model-independent limits on the cross-section for the process
$\ee\rightarrow\h\Zzero$ are derived assuming 100\% decays of the
\h\ into \A\A\ and  100\% decays of the \A\A\ into each of the
following final states: \ccbar\ccbar, \ggbar\ggbar, \ttbar\ttbar,
\ccbar\ggbar, \ggbar\ttbar and \ccbar\ttbar. The results are also
interpreted in the CP-conserving no-mixing MSSM scenario, where
the region 45 $\le \mh \le 85$ \gevocc\ and $2\le \ma \le 9.5$
\gevocc\ is excluded.
\bigskip
\bc
{To be submitted to European Physics Journal C}\\
\ec

\end{titlepage}

\pagenumbering{arabic}

%%%%%%%%%%%%%%%%%%%%%%%%%%%%%%%
% authors list
%%%%%%%%%%%%%%%%%%%%%%%%%%%%%%%
\begin{center}{
%begin authorlist PLEASE DO NOT DELETE THIS COMMENT
G.\thinspace Abbiendi$^{  2}$, C.\thinspace Ainsley$^{  5}$,
P.F.\thinspace {\AA}kesson$^{  3}$, G.\thinspace Alexander$^{
22}$, J.\thinspace Allison$^{ 16}$, P.\thinspace Amaral$^{  9}$,
G.\thinspace Anagnostou$^{  1}$, K.J.\thinspace Anderson$^{  9}$,
S.\thinspace Arcelli$^{  2}$, S.\thinspace Asai$^{ 23}$
% ...
%end authorlist PLEASE DO NOT DELETE THIS COMMENT
}\end{center}\bigskip

\bigskip
%begin institutes
$^{  1}$School of Physics and Astronomy, University of Birmingham,
Birmingham B15 2TT, UK
\\
$^{  2}$Dipartimento di Fisica dell' Universit\`a di Bologna and
INFN, I-40126 Bologna, Italy
\\
$^{  3}$Physikalisches Institut, Universit\"at Bonn, D-53115 Bonn,
Germany
\\
% ...
$^{  5}$Cavendish Laboratory, Cambridge CB3 0HE, UK
\\
% ...
$^{  9}$Enrico Fermi Institute and Department of Physics,
University of Chicago, Chicago IL 60637, USA
\\
% ...
$^{ 16}$Department of Physics, Schuster Laboratory, The
University, Manchester M13 9PL, UK
\\
% ...
$^{ 23}$International Centre for Elementary Particle Physics and
Department of Physics, University of Tokyo, Tokyo 113-0033, and
Kobe University, Kobe 657-8501, Japan
\\
% ...
$^{  e}$ and Department of Experimental Physics, Lajos Kossuth
University,
 Debrecen, Hungary
\\
% ...
%end notes
\newpage
%%%%%%%%%%%%%%%%%%%%%%%%%%%%%%%
\section{Introduction}
%%%%%%%%%%%%%%%%%%%%%%%%%%%%%%%
\label{sec:intro}

The Standard Model (SM) has only one complex doublet of Higgs
fields, resulting in one physical mass eigenstate, the neutral
Higgs scalar boson~\cite{Higgs}. Theoretical limitations of the SM
have prompted the development of many other Higgs models. Possible
extensions of the SM include the Two Higgs Doublet Models (2HDM).
These models predict two complex doublets of scalar fields
resulting in five physical Higgs bosons: two neutral CP-even
scalars, \h\ and \Ho\ (with $\mh < \mH$), one CP-odd scalar, \A,
and two charged scalars, \Hpm~\cite{hhg}. The Higgs sector of the
Minimal Supersymmetric extension of the Standard Model (MSSM) is
of 2HDM(II) type where the introduction of supersymmetry adds new
particles and constrains the parameter space of the model. Due to
the Higgs boson self-coupling, decays of Higgs bosons into other
Higgs bosons become possible if kinematically allowed. In the
2HDM(II), the coupling of the \h\ to \A\A\ is proportional to:
\beq [g~{m_{\mathrm Z}}/{2 \cos(\theta_{\mathrm W})}]
\cos(2\beta)\sin(\beta+\alpha), \eeq where $g$ is the standard
SU(2) gauge coupling, $m_{\mathrm Z}$ the mass of the \Zzero,
$\mathrm{\theta_{\mathrm W}}$ is the weak mixing angle,
$\mathrm{\alpha}$ is the mixing angle that relates the CP-even
Higgs states \Ho\ and \h\ to the field doublets, and
$\mathrm{\tan\beta}$ is the ratio of the vacuum expectation values
of the Higgs scalar fields. When kinematically allowed, the decay
$\h \to \A\A$ may dominate.

The Standard Model predicts that Higgs bosons accessible at LEP
centre-of-mass energies decay preferentially into the heaviest
available fermions since the coupling to the Higgs boson is
proportional to the fermion mass. This fact has motivated the vast
majority of the SM Higgs boson analyses to focus on Higgs boson
decays via b quark and $\tau$ lepton pairs~\cite{adlo}. OPAL has
also performed flavour-independent searches to explore other
possibilities~\cite{2hdm}. Despite this effort, the region with
$\ma < 10$ \gevocc\ and \mh\ between 45 and 86 \gevocc\ remains
uncovered within the 2HDM(II) and MSSM parameter scans. The
analysis described in this paper is dedicated to a narrow Higgs
boson mass region, allowing for the selection of events with very
specific kinematics. This guarantees a higher signal detection
efficiency while achieving an excellent background rejection. This
region has also been investigated by flavour-independent
\cite{2hdm} and decay-mode independent \cite{jochen} analyses but
with lower sensitivity.

The properties of the Higgs bosons in the MSSM can be studied in
the framework of a constrained model with seven parameters:
$M_{\mathrm SUSY}$, $M_2$, $\mu$, $A_{\mathrm \tilde{q}}$,
$\tan\beta$, $\ma$ and $m_{\mathrm \tilde{g}}$. $M_{\mathrm SUSY}$
is the sfermion mass and $M_2$ is the SU(2) gaugino mass
parameter, both at the electroweak scale. The parameter $\mu$ is
the supersymmetric Higgs boson mass parameter, $A_{\mathrm
\tilde{q}}$ is the trilinear Higgs boson-squark coupling
parameter, assumed to be the same for up-type squarks and
down-type squarks, and $m_{\mathrm \tilde{g}}$ is the gluino mass.
As an example, our results are interpreted in the MSSM no-mixing
benchmark scenario~\cite{no-mixing}, which assumes that there is
no mixing between the scalar partners of the left-handed and
right-handed top quarks, and is determined by the following choice
of parameters: $M_{\mathrm SUSY} = 1$ TeV/$c^2$, $M_2 = 200$
\gevocc, $\mu = -200$ GeV, the stop mixing parameter $X_{\mathrm
t} \equiv A_{\mathrm \tilde{q}} - \mu \cot\beta = 0$, $0.4 <
\tan\beta < 50$, 4 \gevocc\ $<$ \ma $<$ 1 TeV/$c^2$ and
$m_{\mathrm \tilde{g}}$ = 800 \gevocc. In this scenario, the
region for $\ma < 10$ \gevocc\ and \mh\ within 45--86 \gevocc\ is
not excluded by the OPAL data, and a smaller mass range with $\ma
< 10$ \gevocc\ and \mh~around 70--86 \gevocc\ also remains
unexcluded in the LEP combined data~\cite{mssmeps01}. In this
region, the \A~boson is too light to decay into \bbbar~ and would
hence have escaped detection by the analyses using b-tagging.
Furthermore, the existing flavour-independent analyses lack the
necessary sensitivity to detect or exclude such possibilities. The
region for \mh~below 45 \gevocc\ is excluded by LEP 1 \cite{2hdm}.
Although this search was originally motivated by the MSSM and 2HDM
studies, our results can be extended to any model that predicts
the existence of at least one scalar and one pseudoscalar Higgs
boson in the mass range of interest and also within models where
the physical Higgs bosons are not CP eigenstates, like the CP
violating MSSM~\cite{CP-MSSM}. For illustrative purposes, we will
use the MSSM no-mixing benchmark scenario as a reference in the
rest of this paper.

The paper is organized as follows: Section \ref{sec:data}
describes the data samples and the OPAL detector. Section
\ref{sec:mc} gives details of the Monte Carlo simulations and
Section \ref{sec:selection} describes the event selection. Section
\ref{sec:systematics} covers systematic uncertainties and the
results are given in Section~\ref{sec:results}.

\section{Data samples and the OPAL detector}
\label{sec:data} The data used for this analysis were collected
during 1998--2000 at LEP in \ee~collisions at centre-of-mass
energies (\ecm) between 188 and 209 GeV. The data sample is
divided into four subsamples, namely 188 $\le$ \ecm$ \le$ 193 GeV
(201.7 \pb), 193 $\le$ \ecm$ \le$ 198 (75.1 \pb), 198 $ \le$ \ecm$
\le$ 203.5 GeV (115.2 \pb) and 203.5 $<$ \ecm$ \le$ 209 GeV (206.7
\pb). The analysis is performed separately on each sub-sample and
the results are combined. The choice of the subgrouping of
centre-of-mass energies is justified by the fact that the
production cross-section for the signal does not change
appreciably within each of the four subsamples since, for this
analysis, we study a region far below the kinematic limit.

The OPAL detector~\cite{detector} has nearly complete solid angle
coverage and hermeticity. The innermost detector of the central
tracking is a high-resolution silicon microstrip vertex
detector~\cite{simvtx} which lies immediately outside of the beam
pipe.  Its coverage in polar angle\footnote{OPAL uses a
right-handed coordinate system where the $+z$ direction is along
the electron beam and where $+x$ points to the centre of the LEP
ring. The polar angle, $\theta$, is defined with respect to the
$+z$ direction and the azimuthal angle, $\phi$, with respect to
the horizontal, $+x$ direction.} is $|\cos\theta|<0.93$.   The
silicon microvertex detector is surrounded by a high precision
vertex drift chamber, a large volume jet chamber, and $z$-chambers
to measure the $z$ coordinates of tracks, all in a uniform 0.435~T
axial magnetic field. The lead-glass electromagnetic calorimeter
and the presampler are located outside the magnet coil. In
combination with the forward detectors, namely the forward
calorimeters, a forward ring of lead-scintillator modules (the
``gamma catcher'')~\cite{detector}, a forward scintillating tile
counter (the ``MIP plug'')~\cite{mipplug}, and the
silicon-tungsten luminometer~\cite{sw}, the calorimeters provide a
geometrical acceptance down to 24~mrad from the beam direction.
The silicon-tungsten luminometer measures the integrated
luminosity using Bhabha scattering at small angles~\cite{lumino}.
The magnet return yoke is instrumented with streamer tubes and
thin gap chambers for hadron calorimetry and is surrounded by
several layers of muon chambers.

Events are reconstructed from charged particle tracks and
energy deposits (``clusters") in the electromagnetic and hadron calorimeters.
The tracks and clusters must pass a set of quality requirements
similar to those used in
previous OPAL Higgs boson searches~\cite{higgsold}.
In calculating the total visible energies and momenta, $E_{\mathrm vis}$
and $\vec{P}_{\mathrm vis}$, of events and
individual jets~\cite{drm}, corrections are applied to prevent the
double counting of energy of tracks with associated
clusters~\cite{lep2neutralino}.

\section{Monte Carlo simulation}
\label{sec:mc}

Monte Carlo samples for signal and background were generated at
four different centre-of-mass energies, namely 189, 196, 200 and
206 GeV, chosen to be close to the mean centre-of-mass energy in
each data subsample described in Section~\ref{sec:data}.

\begin{figure}[htb]
 \begin{center}
   \mbox{
    \epsfysize=4.0cm
    \epsffile{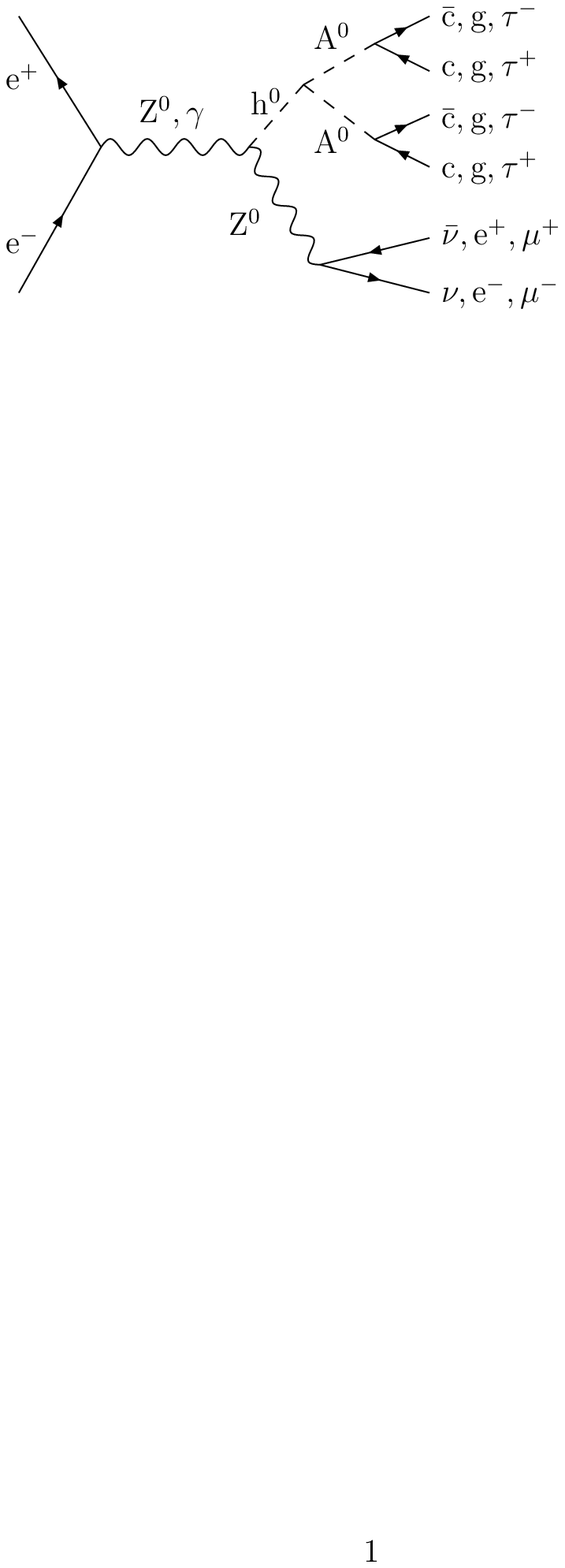}
   }
 \parbox{0.9\textwidth}{\caption {\sl
The Feynman diagram for the processes considered in this analysis.
\label{feynman}}}
\end{center}
\end{figure}

We study only $\h\Zzero$ production since, in the parameter space
region of interest for our analysis, its cross-section is about
ten times larger than that for $\h\A$ production in the MSSM. The
\h~is forced to decay into two \A~bosons, $\h \to \A\A$, and each
\A~can decay into any of the following channels: \ccbar,
\ttbar~and \ggbar. Resonances are not included in the simulation
of \A\ decays. For example, in the MSSM no-mixing scenario, for
$3.3~{\mathrm GeV/c^2} < \ma < 9.5$ \gevocc, the \A\ branching
fractions into \ccbar\ and \ttbar\ are 0.5-0.9 and 0.4-0.05,
depending on the value of \tanb. Below the \ttbar~threshold, the
\A~decays nearly exclusively into a gluon pair. Two different
\Zzero~decay modes are investigated: $\Zzero \to \nnbar$ and
$\Zzero \to \llbar$ with $\ell=$e or $\mu$. For each of the
\Zzero~decay modes, the six final states obtained by all possible
combinations of the \A\ decays to gg, \ccbar~and \ttbar\ have been
analysed. In the no-mixing MSSM scenario below the production
threshold for \bbbar, these final states account for between 75\%
and 100\% of the total decays of the \A\ boson~\cite{feynhiggs}.
The corresponding Feynman diagram is given in Figure
\ref{feynman}.

Monte Carlo samples were generated with \ma=2, 4, 6, 9 and 11
\gevocc\ and for \mh~= 45, 60, 70, 80, 86 \gevocc\ at each of the
four centre-of-mass energies considered. For each [$\ma,~\mh$]
combination and each \Zzero~decay channel studied, we produced
3000 events for each of the six possible final states using the
HZHA2~\cite{HZHA} generator and the full OPAL detector
simulation~\cite{gopal}.

The branching fraction BR$(\h \to \A\A)$ is relatively constant
for \ma~in the range of 1 to 11 \gevocc\ for a given value of \mh.
The $\ee \to \h\Zzero$ production cross-section does not depend
strongly on \mh~in the range $45\le\mh\le 86$~\gevocc\ but
increases with increasing $\tan\beta$ values.

Monte Carlo simulations are also used to study the various \sm\
background processes. The 2-fermion events, $\ee \to \qqbar$, are
simulated with the KK2f generator using CEEX~\cite{CEEX} for the
modelling of the initial state radiation and PYTHIA
6.125~\cite{pythia} for the fragmentation and hadronisation
processes. Bhabha events are generated with the
BHWIDE~\cite{bhwide} and TEEGG~\cite{teegg} generators,
$\ee\rightarrow\mm$ and $\ee\rightarrow\tau^+\tau^-$ events are
simulated with the KK2f generator using CEEX and ISR-FSR
interference. The 4-fermion samples are generated with
grc4f~\cite{grc4f} for the $\ell\ell\ell\ell$, $\ell\ell \qqbar$,
$\ell\nu \qqbar$, $\nu\nu \qqbar$ and $\qqbar\qqbar$ channels,
where $\ell=$ e, $\mu, \tau$. One 2-photon sample generated at 200
GeV is used for centre-of-mass energies of 189, 196 and 200 GeV
and the resulting cross-sections normalised according to the
centre-of-mass energy. An independent sample is used at 206 GeV.
The various 2-photon processes are modelled by the
Vermaseren~\cite{Vermaseren}, HERWIG~\cite{Herwig},
Phojet~\cite{Phojet} and F2GEN~\cite{f2gen} generators. Typically,
at each centre-of-mass energy, the generated 4-fermion and
2-fermion samples are 30 times larger than the data sample, and
the 2-photon sample is two to ten times larger than the data.

\section{Event selection}
\label{sec:selection}  In the low \A\ mass region covered by this
search the separation between the decay products of the \A\ tends
to be small, and they are generally reconstructed as a single jet.
The final event topology consists of two jets recoiling against
the \Zzero~decay products. The invariant mass of a single jet
reproduces the mass of the \A\ while the mass of the combined
two-jet system reproduces the mass of the \h. A mass resolution
between 0.5 and 3.0 \gevocc\ is achieved for \ma, while for
\mh~the resolution is about 8--16 \gevocc, all depending on the
\Zzero~decay mode and the Higgs boson masses.

We perform three separate analyses based on three different
\Zzero\ decay channels: $\Zzero \to \nnbar$, \mm\ and \ee. The
event selection criteria are described in detail in the next two
sections. After this initial selection, a likelihood variable is
constructed for each channel to increase the sensitivity. This
likelihood variable is used to perform a scan and set limits.

\subsection{The \boldmath $\Zzero \to \nnbar$ channel \unboldmath
} \label{sec:znn} Events in the process $\h\Zzero \to \A\A \nnbar$
are characterised by two jets recoiling against an invisible
\Zzero. Each event is forced into a two-jet topology using the
Durham algorithm~\cite{drm} and a 1-constraint kinematic fit is
performed, requiring energy and momentum conservation and forcing
the missing mass to be equal to $m_{\mathrm Z}$. Background from
cosmic ray events is removed by requiring at least one track per
event and applying the two cosmic-ray vetoes described
in~\cite{iellco} and~\cite{ieggt0}. The visible invariant mass is
also required to exceed 5 \gevocc\ to match a corresponding cut
used in the generation for the 2-photon MC samples. This cut has
no effect on this analysis since it is far from the region of
interest for the signal.

The selection criteria used to search for $\h\Zzero\to\A\A\nnbar$
are mainly based on event shape variables and reconstructed
masses. The first four preselection cuts guarantee general data
quality, confinement within the detector region and rejection of
2-photon background. The other more specific selection criteria
aim mostly at rejecting the other backgrounds. The cuts used are
described below:

\begin{enumerate}
\item
  The fraction of total visible energy in the forward
  detectors must be less than 60\%,
  the event transverse momentum measured w.r.t. the beam axis must exceed 3 GeV and
  the visible energy of the event must exceed 20\% of the centre-of-mass energy.

\item Events are rejected if they have energy deposits exceeding 2 GeV, 5
GeV and 5 GeV in the forward calorimeter, the silicon-tungsten
luminometer and the gamma catcher, respectively. This requirement
rejects events with initial state radiation or particles escaping
detection in the beam pipe.

\item The fraction of good tracks as defined in~\cite{higgsold} relative to
the total number of tracks should be greater than 0.2.

\item The polar angle, $\theta$, of each jet must satisfy $|\cos\theta| <
0.9$ to reject events with jets partially contained in the
detector. This is also a very powerful cut to reduce background
contributions from 2-photon and $\mathrm{Z}/\gamma^*$ events with
large initial state radiation.

\item
The invariant mass of the di-jet system after the kinematic fit
must be in the range 30--95 \gevocc. This mass should reproduce
\mh\ except in the case of $\A \to \tau^+\tau^-$ where there are
missing neutrinos.

\item The aplanarity\footnote{The aplanarity is defined as
${3\over 2}\lambda_3$, where $\lambda_i$ are the eigenvalues
[$\lambda_1\ge\lambda_2\ge\lambda_3$ with
$\lambda_1+\lambda_2+\lambda_3=1$] of the sphericity tensor
$S^{\alpha\beta}={\sum_{i}p_i^\alpha p_i^\beta/\sum_{i}|{\bf
p}_i|^2}$, and is related to the transverse momentum component out
of the event plane.} must be in the range 0.0002--0.03 and the
event shape variable $C$~\cite{cparam} must be less than 0.8.

\item The polar angle of the missing
momentum vector must satisfy $|\cos\theta| < 0.97.$

\item The invariant mass of the more energetic jet
must be between 0.5 and 13.0 \gevocc. (The mass of the
reconstructed jet corresponds to \ma~within the detector
resolution. Since we perform a mass-independent search, we allow
for a broad range of values for \ma.)

\item The invariant mass of the less energetic jet
must be between 0.5 and 10.5 \gevocc.

\end{enumerate}

The numbers of events passing cuts 1 to 9 are shown in Table
\ref{cut300} for all centre-of-mass energies combined. The numbers
of events found in the data are compared to the various
backgrounds after each cut. The \qqbar\ and $\ell^+\ell^-$
contributions to the 2-fermion background are listed in two
separate columns. The numbers of events expected for a signal
hypothesis of \ma~=~6 \gevocc\ and \mh~=~70 \gevocc\ in the MSSM
no-mixing scenario and the corresponding efficiencies are also
shown.

\begin{table}[htb]
\begin{center}
\begin{tabular}{|c|c|c|c|c|c|c|c|} \hline
\multicolumn{8}{|c|}{$\Zzero \to \nnbar$ channel for \ecm=189--209
GeV combined}
\\\hline cut& data &  total& \multicolumn{4}{|c|}{background
sources}&signal\\\cline{4-7} && bgnd &\qqbar  & $\ell^+\ell^-$ &
4f & \twog &  (efficiency) \\\hline
 $1$ &  21584 & 20733.2 &  9879.4 &  4265.1&  2522.5 &  4066.3  & 73.7 (93.2\%) \\
 $2$ &  13241 & 12776.5 &  7793.0 &  1688.2&  2044.5 &  1250.8  & 69.2 (87.6\%) \\
 $3$ &  12990 & 12597.8 &  7714.4 &  1644.2&  2012.9 &  1226.4  & 68.8 (87.1\%) \\
 $4$ &   9015 &  8876.9 &  5281.1 &  1300.5&  1617.0 &   678.4  & 68.8 (87.1\%) \\
 $5$ &   7837 &  7809.4 &  4966.2 &  1003.1&  1332.5 &   507.6  & 59.4 (75.2\%) \\
 $6$ &   5227 &  5338.8 &  3991.0 &   206.0&   776.2 &   365.6  & 50.9 (64.4\%) \\
 $7$ &   1571 &  1547.5 &   731.8 &   142.3&   585.0 &    88.4  & 50.0 (63.3\%) \\
 $8$ &    854 &   839.8 &   429.1 &   111.1&   237.8 &    61.8  & 48.6 (61.5\%) \\
 $9$ &    475 &   470.5 &   271.6 &    57.7&   101.6 &    39.6  & 47.7 (60.4\%) \\ \hline

${\cal L} > 0.88$&    18 &    14.9 &     0.1 &     1.0&    13.8 &
0.0 & 38.0 (48.1\%) \\ \hline
\end{tabular}
\parbox{0.9\textwidth}{\caption {\sl
The numbers of observed events together with the Monte Carlo
expectation for the various sources of background given for the
combined sample (189 $ \le E_{\mathrm CM}\le 209$ GeV) for the
$\Zzero \to \nnbar$ channel. The corresponding efficiencies are
given within parentheses for one signal hypothesis (\ma~= 6
\gevocc, \mh\ = 70 \gevocc) in the MSSM no-mixing scenario. The
various cuts are described in the text. The likelihood cut is only
given for illustrative purposes. It is not used to set the limits
in Section \ref{sec:results}. \label{cut300}}}
\end{center}
\end{table}

\subsubsection{Likelihood selection} \label{sec:likelihood} 
\begin{figure}[htb]
 \begin{center}
   \mbox{
    \epsfxsize=\textwidth
    \epsffile{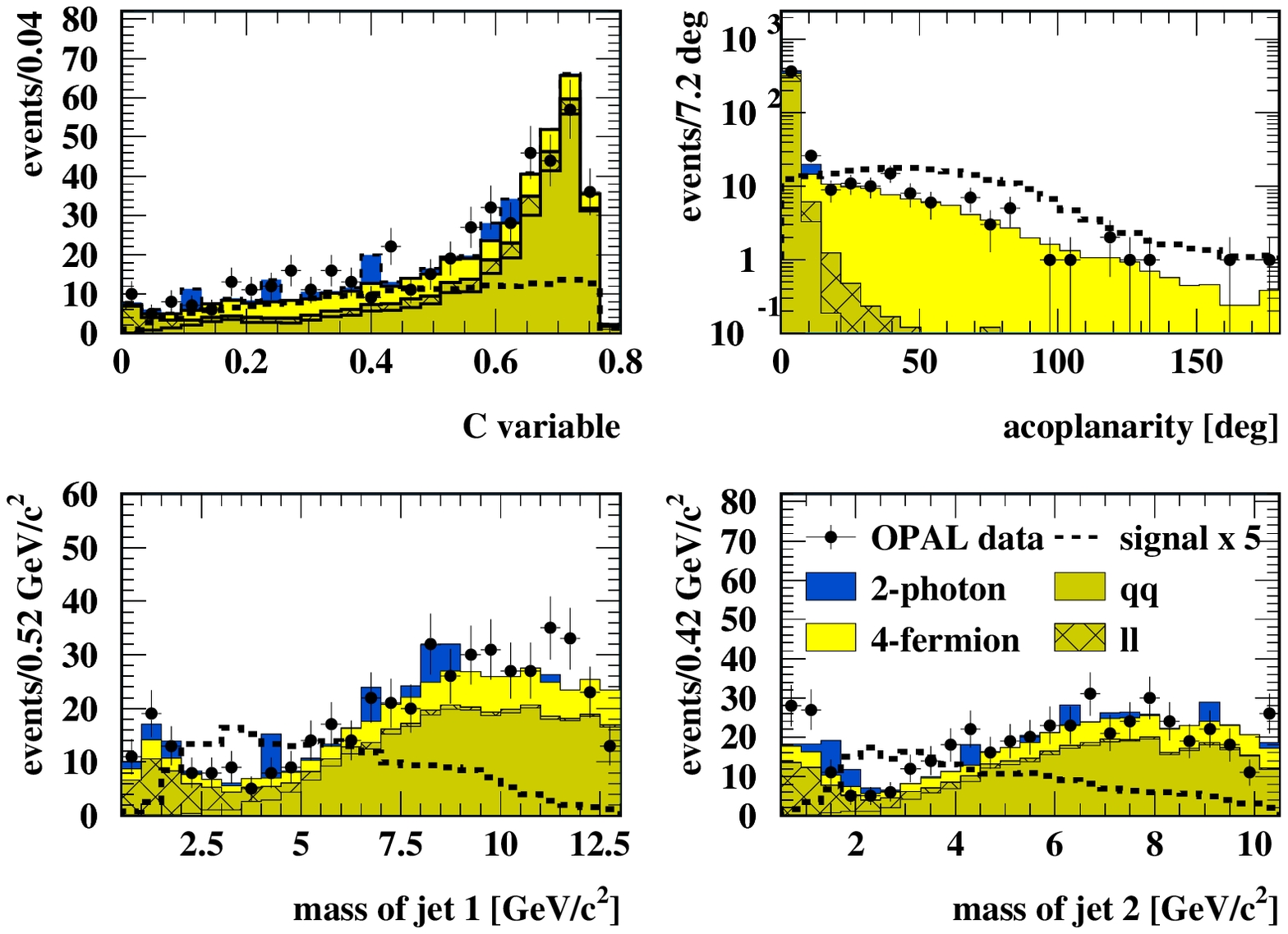}
   }
 \parbox{0.9\textwidth}{\caption {\sl
The four input variables used for the likelihood function of the
$\Zzero \to \nnbar$ channel after all preselection cuts are
applied. The contributions from the Standard Model backgrounds are
added and normalised to the data integrated luminosity. The
results are shown here for \ecm=189--209 GeV combined. The
contribution from the reference signal is scaled up by a factor of
five.\label{after300}}}
\end{center}
\end{figure}

A
discriminating variable is formed by combining information from
the following four variables into a likelihood based on the
Projections and Correlations Method~\cite{pc104}:
\begin{enumerate}
\item The event shape variable $C$.
\item The acoplanarity\footnote{The acoplanarity angle
is the absolute value of 180$^\circ$ minus the opening angle
between the two jets in the plane transverse to the beam
direction.} angle of the two jets.
\item The invariant mass of the more energetic jet.
\item The invariant mass of the less energetic jet.
\end{enumerate}
\vspace{-0.2cm} These variables are shown for the data, the
different sources of background and the reference signal in Figure
\ref{after300} after all preselection cuts given in Section
\ref{sec:znn} are applied and for all centre-of-mass energies
combined. To form the reference distributions for the signal,
%used by the likelihood function,
the distributions for sixteen [\ma,\mh] mass hypotheses obtained
for \mh = 60, 70, 80, 86 \gevocc\ and \ma= 2, 4, 6, 9 \gevocc\ are
summed after relative re-normalization  according to the
integrated luminosity of the data and their production
cross-section. For the background, three reference distributions
are used: the $\ell^+\ell^-$, the 4-fermion, and the combined
2-photon and \qqbar\ samples. Hence, the data are compared to four
reference distributions: three for the \sm~background and one
single distribution for the signal. The likelihood function used
to derive our result is formed separately for each centre-of-mass
energy considered. The distribution of the likelihood input
variables for all centre-of-mass energies combined is shown in
Figure \ref{after300} for illustrative purposes only. The
efficiencies for each mass hypothesis are also determined
separately at each of the [\ma,\mh] mass hypotheses and for each
centre-of-mass energy. A small correction has to be applied to the
efficiencies and backgrounds due to accelerator-related
backgrounds in the forward detectors which are not simulated. From
random beam crossing events the correction factors have been
evaluated to be 3.1$\%$, 3.6$\%$, 3.6$\%$ and 3.2$\%$ for
$\sqrt{s}$ = 189, 196, 200 and 206 GeV, respectively.

For illustrative purposes, a cut
at 0.88 placed on the likelihood variable would reject most
background events and retain sufficient efficiency at all mass
hypotheses considered. The same cut would be used for all data
sets and is chosen by maximising the signal purity times the
efficiency of our signal reference at all energies. After this
cut, 18 events would be retained in data compared to $14.9$
expected from SM backgrounds. The likelihood distribution function
is shown in Figure \ref{like300} (a) for the data, the
\sm~backgrounds and the reference signal for the 189--209 GeV data
combined. The likelihood cut is not used to set limits in Section
\ref{sec:results}. The signal efficiency ranges from 38\% to 75\%
for $\ma = 6$ \gevocc (see Figure \ref{eff}(a)). It drops down
between 21\% and 53\% for $\ma = 11$ \gevocc.

\subsection{The  \boldmath $\Zzero \to \mm$ and $\Zzero \to \ee$
channels \unboldmath } \label{sec:leptonic} The leptonic
\Zzero~channels, namely $ \h\Zzero \to\A\A \ee$ and $ \h\Zzero
\to\A\A \mm$ are also investigated. These events are characterised
by the presence of two jets with invariant mass compatible with
the \h~mass and a lepton pair with invariant mass close to the
\Zzero~mass.

The electron and muon analyses share the first three preselection
cuts listed below. The selection starts with lepton
identification. First the event is required to have one isolated
lepton in association with two jets by applying the same selection
criteria used in~\cite{ww172} for the isolation and identification
of a lepton in qq$\ell\nu$ events from \WW~decays. The selection
is based on the probability for a track to be correctly associated
with an isolated lepton. The probability is obtained with a
likelihood method based on kinematic and lepton identification
variables. No requirement on the number of tracks is made to avoid
biasing the selection against low multiplicity events. Then the
identification of two isolated leptons with the same flavour and
opposite charge produced in association with two jets is required
and the selected events are forced into a two jet configuration
using the Durham algorithm without including the two best lepton
candidates.

The next two cuts are applied to ensure confinement within the
detector while the remaining selection criteria are optimised for
background rejection and differ for the two analyses. This was
necessary in order to reduce the large 2-photon background
contribution in the electron channel. All cuts are optimised to
maximise purity times efficiency for a mixture of all signal
hypotheses.

\begin{itemize}
\item[1.] The isolation and identification of two oppositely charged leptons (\ee\ or \mm) in association with
two jets.

\item[2.] To avoid events having particles lost in the beampipe, both jets
    must have $|\cos \theta| < 0.99$.

\item[3.] The visible energy must be greater than $0.78$ of the
centre-of-mass energy. This cut ensures the event is
well-contained within the detector and rejects some of the
4-fermion background.
\end{itemize}

{\bf{Muon channel}}

\begin{itemize}

\item [4a.] The invariant mass of the more energetic jet is required to
    be less than 25 \gevocc.

\item [5a.] The invariant mass of the less energetic jet
    has to be less than 15 \gevocc.

\end{itemize}

{\bf{Electron channel}}

\begin{itemize}

\item [4b.] The invariant mass of the two leptons should be between
           66 and 115 \gevocc.

\item [5b.] The number of charged tracks in each jet should be less than 10.
          This cut drastically reduces the 4-fermion background.

\item [6b.] The invariant mass of the more energetic jet is required to
    be less than 36 \gevocc.

\item [7b.] The invariant mass of the less energetic jet must be less than
    30 \gevocc.

\item [8b.] The angle between the two jets must exceed 1.6 rad to
    reduce the 2-photon background.

\end{itemize}

\begin{table}[htb]
\begin{center}
\begin{tabular}{|c|c|c|c|c|c|c|}\hline
\multicolumn{7}{|c|}{muon channel for \ecm=189--209 GeV combined}
\\\hline
cut& data &  total& \multicolumn{3}{|c|}{background
sources}&signal\\\cline{4-6}
&&bgnd& 2f  &   4f & \twog  & (efficiency) \\
\hline
1            &   56 &   61.0      & 0.7  & 60.0   & 0.1      & 10.1 (77.4\%) \\
2            &   55 &   60.0      & 0.6  & 59.4   & 0.0      & 10.1 (77.4\%) \\
3            &   44 &   47.4      & 0.2  & 47.2   & 0.0      &  9.5 (72.8\%)\\
4a           &   33 &   37.6      & 0.0  & 37.6   & 0.0      &  9.1 (69.8\%)\\
5a           &   27 &   30.5      & 0.0  & 30.5   & 0.0      &  8.8 (67.5\%)\\
\hline $\cal L$ $ > 0.56$ &    4 & $3.6$ & 0.0  & 3.6    & 0.0     &  7.8 (59.8\%)\\ \hline
\end{tabular}
\parbox{0.9\textwidth}{\caption {\sl
The numbers of observed events together with the Monte Carlo
expectation for the various sources of background given for the
combined sample (189 $ \le E_{\mathrm CM}\le 209$ GeV) for the
muon channel. The corresponding efficiencies are given within
parentheses for one signal hypothesis (\ma~= 6 \gevocc, \mh\ = 70
\gevocc) in the MSSM no-mixing scenario. The various cuts are
described in the text. The 2-fermion sample contains both \qqbar\
and \llbar\ events. The likelihood cut is only given for
illustrative purposes and is not used to set the limits in Section
\ref{sec:results}. \label{cutflowm}}}
\end{center}
\end{table}

\begin{table}[htb]
\begin{center}
\begin{tabular}{|c|c|c|c|c|c|c|}\hline
\multicolumn{7}{|c|}{electron channel for \ecm=189--209 GeV
combined}\\\hline cut& data &  total&
\multicolumn{3}{|c|}{background sources}&signal\\\cline{4-6}
&&bgnd& 2f & 4f & \twog & (efficiency)
\\\hline
1   &  100 &  103.6       &  1.9 &   80.4 & 21.1   &  9.9 (75.9\%) \\
2   &   99 &   99.5       &  1.8 &   79.3 & 18.5   &  9.7 (74.4\%) \\
3   &   77 &   79.5       &  1.4 &   62.0 & 16.1   &  9.2 (70.6\%) \\
4b  &   35 &   35.9       &  0.4 &   29.5 &  6.1   &  7.4 (56.7\%) \\
5b  &   23 &   19.7       &  0.2 &   15.2 &  4.3   &  7.1 (54.4\%) \\
6b  &   21 &   17.8       &  0.1 &   14.1 &  3.5   &  6.3 (48.3\%) \\
7b  &   20 &   16.2       &  0.1 &   13.1 &  2.8   &  6.1 (46.8\%) \\
8b  &   19 &   14.3       &  0.1 &   12.3 & 1.8 &6.0(46.0\%)\\
\hline $\cal L$ $ > 0.52$     &    4 & 3.6  & 0.0 & 2.5 &  1.1
&5.1 (39.1\%) \\ \hline
\end{tabular}
\parbox{0.9\textwidth}{\caption {\sl
The numbers of observed events together with the Monte Carlo
expectation for the various sources of background given for the
combined sample (189 $ \le E_{\mathrm CM}\le 209$ GeV) for the
electron channel. The corresponding efficiencies are given within
parentheses for one signal hypothesis (\ma~= 6 \gevocc, \mh\ = 70
\gevocc) in the MSSM no-mixing scenario. The various cuts are
described in the text. The 2-fermion sample contains both \qqbar\
and \llbar\ events. The likelihood cut is only given for
illustrative purposes and is not used to set the limits in Section
\ref{sec:results}. \label{cutflowe}}}
\end{center}
\end{table}

The number of events passing each of these cuts can be found in
Tables \ref{cutflowm} and \ref{cutflowe} for the muon and electron
channels, respectively. The numbers of events selected in the data
are compared with the total background expected from the
4-fermion, 2-fermion and 2-photon samples after each cut. The
number of events expected for a signal hypothesis of \mh~= 70
\gevocc\ and \ma~= 6 \gevocc\ in the MSSM no-mixing scenario is
also shown. The 2-fermion background consists of \qqbar\ and
$\tau^+\tau^-$ events, but only \qqbar\ events survive the
preselection. While in the $\Zzero \rightarrow \mm$~ channel the
two-photon background is fully rejected by the preselection, in
the $\Zzero \rightarrow \ee$ channel some hadronic tagged
two-photon events\footnote{Two-photon events in which one or both
scattered electrons are detected~\cite{yellow2}.} survive the
preselection cuts. To decrease the large statistical error on this
background, all Monte Carlo generated 2-photon events have been
used at each centre-of-mass energy.

\subsubsection{Likelihood selection} \label{sec:likel}
A discriminating variable is formed as for the $\Zzero \to \nnbar$
channel by combining information from the variables listed below
into a likelihood based on the Projections and Correlations
method~\cite{pc104}. The sixteen signal hypotheses obtained for
\mh = 60, 70, 80, 86 \gevocc\ and \ma= 2, 4, 6, 9 \gevocc\ are
combined to form one single reference signal distribution for each
input variable as described in Section \ref{sec:likelihood}. This
is done separately for the two leptonic channels. The variables
used as inputs for the two likelihood functions are described
here:
\newpage
{\bf{Muon channel}}
\begin{enumerate}
\item  Angle between the more energetic muon and the nearest jet.
\item  Angle between the less energetic muon and the nearest jet.
\item  Reconstructed invariant mass of the more energetic jet.
\item  Reconstructed invariant mass of the less energetic jet.
\end{enumerate}
{\bf{Electron channel}}
\begin{enumerate}
\item  Invariant mass of the electron pair.
\item  Angle between the less energetic electron and the nearest jet.
\item  Reconstructed invariant mass of the more energetic jet.
\item  Reconstructed invariant mass of the less energetic jet.
\item  Angle between the two jets.
\end{enumerate}

\begin{figure}[htb]
 \begin{center}
   \mbox{
    \epsfxsize=\textwidth
    \epsffile{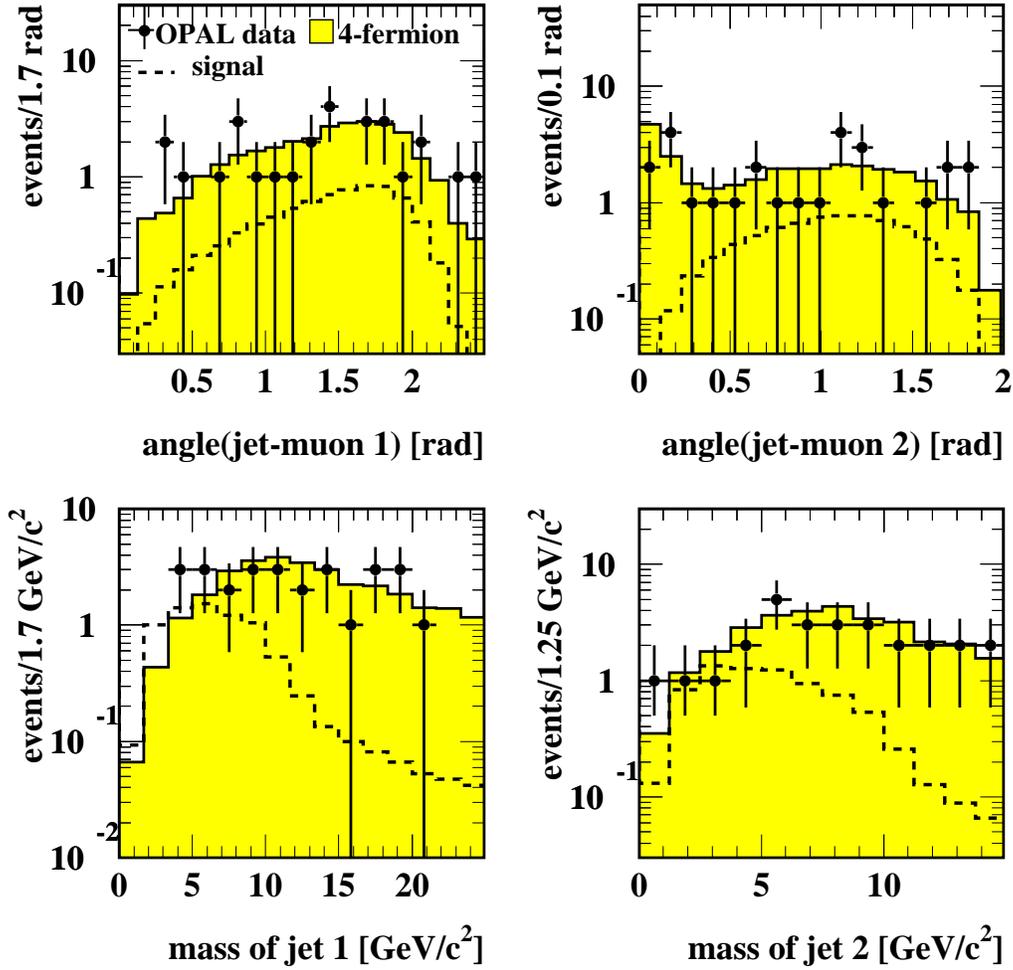}
   }
 \parbox{0.9\textwidth}{\caption {\sl
The four input variables used for the likelihood function in the
muon channel after all preselection cuts  for \ecm=189--209
\gevocc\ combined, where the labels $1$ and $2$ refer to the more
and less energetic muon and jet, respectively. The only
contribution from Standard Model backgrounds surviving the
preselection, namely the 4-fermion sample, is compared to the
data. The contribution from the reference signal is also shown.
\label{aftmu}}}
\end{center}
\end{figure}

\begin{figure}[htb]
 \begin{center}
   \mbox{
    \epsfxsize=\textwidth
    \epsffile{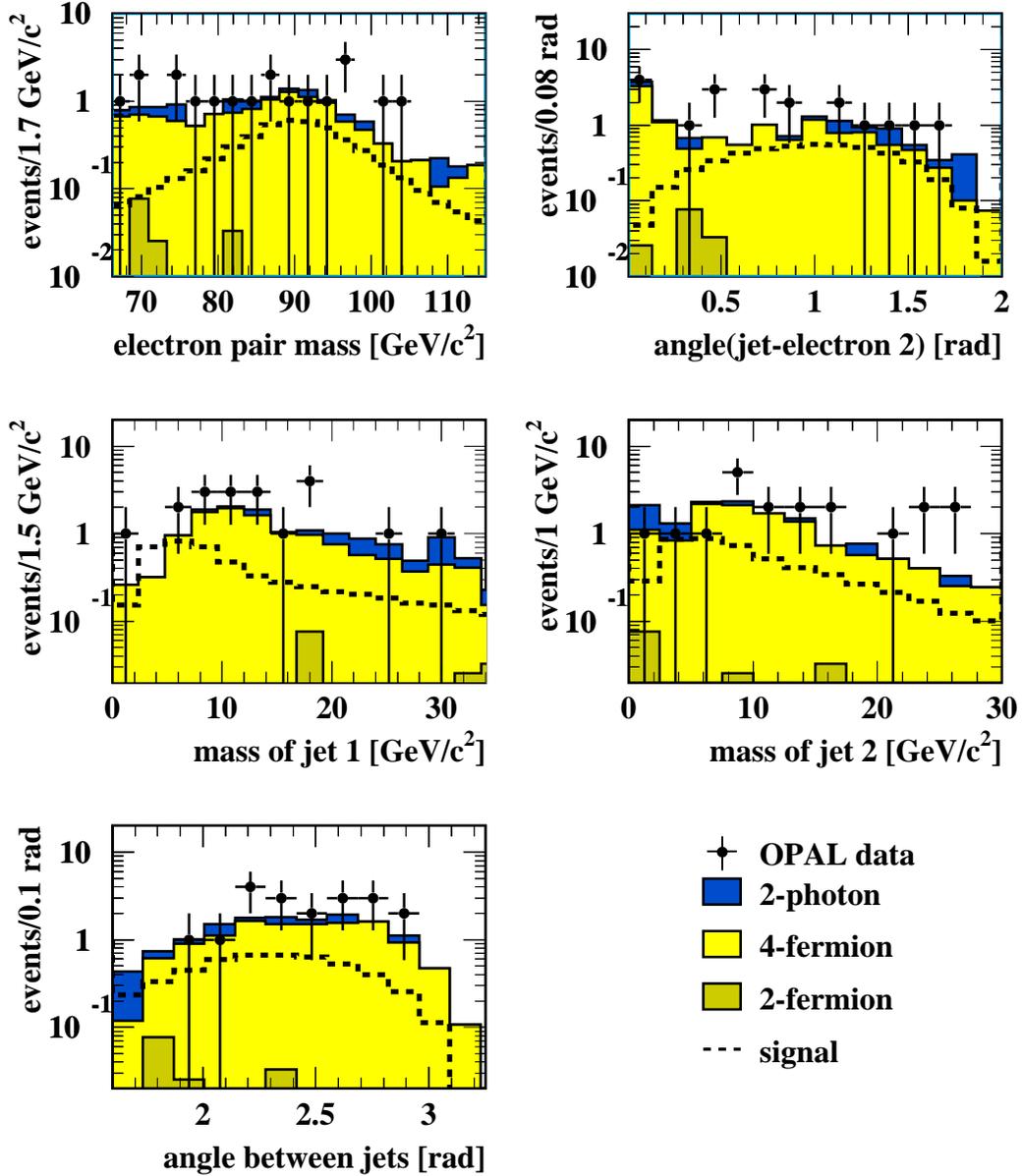}
   }
 \parbox{0.9\textwidth}{\caption {\sl
The five input variables used for the likelihood function after
all preselection cuts in the electron channel for \ecm=189--209
GeV combined, where the labels $1$ and $2$ refer to the more and
less energetic electron and jet, respectively. The contributions
of Standard Model backgrounds surviving the preselection, namely
the 4-fermion, 2-fermion and 2-photon samples, are compared to the
data. The contribution from the reference signal is also
shown.\label{aftel}}}
\end{center}
\end{figure}

The distributions of the input variables for the data, the total
background and the reference signal after applying all
preselection cuts are shown in Figures~\ref{aftmu} and \ref{aftel}
for the muon and electron channels, respectively. %For the muon
For the background, only one reference distribution is used in
both channels consisting of the 4-fermion background sample: too
few 2-photon and 2-fermion events survive the preselection in the
electron channel to use them as reference histograms. The
likelihood distribution function is shown for the data, the
Standard Model backgrounds and the reference signal in
Figures~\ref{like300} (b) and (c) for the muon and electron
analyses, respectively.

\begin{figure}[htb]
  \begin{center}
    \mbox{
     \epsfxsize=\textwidth
     \epsffile{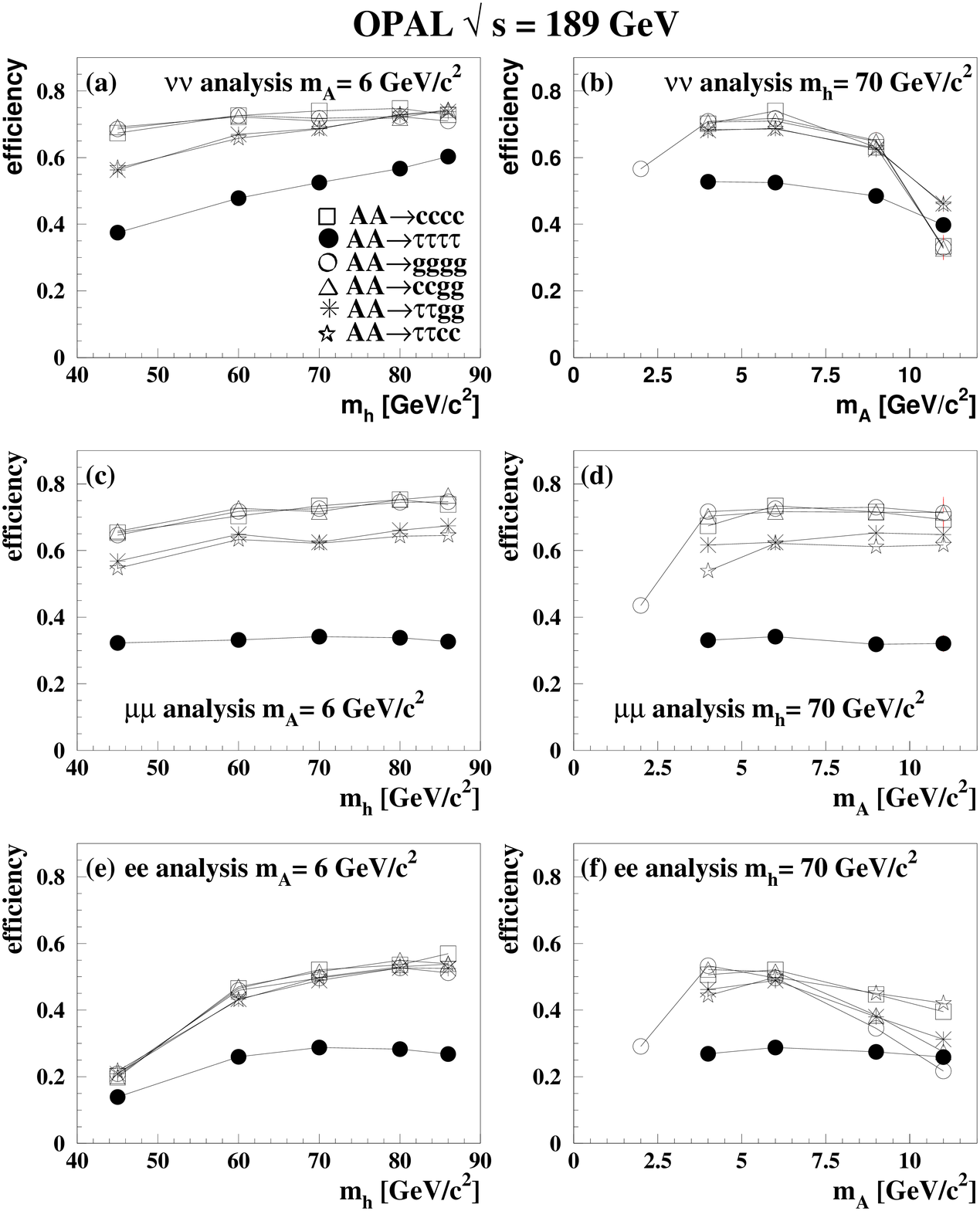}
    }
  \parbox{0.9\textwidth}{\caption {\sl
Signal selection efficiencies at \ecm\ = 189 GeV  versus \mh (\ma)
for \ma\ = 6 \gevocc\ (\mh\ = 70 \gevocc) for the missing energy
and the leptonic channels. The efficiencies are shown without any
cut on the likelihood variable for all six decay channels of a
\A\A~pair with \A~decaying into \ccbar, \ttbar~or \ggbar.
\label{eff}}}
 \end{center}
 \end{figure}

For illustrative purposes, a cut on the likelihood value could be
set at 0.56 for the muon and 0.52 for the electron analysis to
optimise background rejection and signal detection efficiency at
all mass hypotheses considered by maximising the purity times
signal efficiency of our signal reference. This cut would retain
four events in data compared to $3.6$ expected from SM backgrounds
both in the muon and electron channels. This cut is not used to
set limits in Section \ref{sec:results}. For the muon channel, the
signal efficiency ranges from 32\% to 77\% for $\ma = 6$ \gevocc\
(see Figure \ref{eff}(c)) and between 29\% and 75\% for $\ma = 11$
\gevocc. For the electron channel, the signal efficiency ranges
between 14 and 57\% at $\ma = 6$ \gevocc\ (see Figure
\ref{eff}(e)) and between 4 and 46\% at $\ma = 11$ \gevocc.

\section{Systematic uncertainties}
\label{sec:systematics} Many effects related to possible
inadequacies in the simulation of physical quantities in the Monte
Carlo samples contribute to the systematic uncertainty. These
contributions are listed in Table \ref{syst} for the $\Zzero \to
\nnbar$, $\Zzero \to \mu^+ \mu^-$ and $\Zzero \to \ee$ channels.
They are added in quadrature to obtain the total systematic
uncertainty for each channel. The evaluation of each source
considered is described here:

\begin{table}[tb]
\begin{center}
\begin{tabular}{|c|c|c|c|c|c|c|} \hline
\multicolumn{1}{|c|}{} & \multicolumn{2}{|c|}{$\Zzero \to \nnbar$
channel} & \multicolumn{2}{|c|}{$\Zzero \to \mu^+ \mu^-$ channel}
& \multicolumn{2}{|c|}{$\Zzero \to \ee$ channel} \\ \hline source
& bgnd   & signal        & bgnd     & signal & bgnd     & signal
\\ \hline
$\tan\lambda$      &$\pm0.1$&$\pm(0.0-1.3)$ & $\pm0.3$ & $\pm(0.0-1.6)$ & $\pm0.4$ & $\pm(0.0-0.7)$ \\
$\kappa$           &$\pm0.2$&$\pm(0.1-1.1)$ & $\pm1.2$ & $\pm(0.2-7.0)$ & $\pm2.7$ & $\pm(0.3-2.2)$ \\
inputs to ${\cal L}$&$\pm8.0$&$\pm(0.7-8.3)$& $\pm8.9$ & $\pm(1.0-4.2)$ & $\pm10.2$& $\pm(2.0-8.3)$ \\
SM cross-sec.      &$\pm2.0$&   --           & $\pm2.0$ &     --          & $\pm6.7$ &      --       \\
MC statistics      &$\pm4.8$&$\pm(1.4-13.5)$& $\pm10.0$& $\pm(1.4-11.9)$& $\pm11.7$& $\pm(2.0-10.5)$ \\
lepton ID          &   --    &   --           & $\pm0.9$ &
$\pm0.9$ & $\pm2.3$ & $\pm2.3    $ \\ \hline total
&$\pm9.8$&$\pm(1.7-15.9)$& $\pm13.6$& $\pm(2.2-14.5)$& $\pm17.3$&
$\pm(4.4-13.7) $ \\ \hline

\end{tabular}
\parbox{0.9\textwidth}{\caption {\sl
Contributions to the systematic uncertainties for the $\Zzero \to
\nnbar$, $\Zzero \to \mm$ and $\Zzero \to \ee$ channels expressed
in percent for the signal Monte Carlo and the Standard Model
background, as described in the text. The error on the signal
corresponds to the range of values obtained for the generated
[\ma,\mh] values for MSSM Higgs bosons in the no-mixing scenario.
\label{syst}}}
\end{center}
\end{table}

\bi
\item {\bf Simulation of likelihood input variables}.
Each likelihood input variable is re-scaled in the Monte Carlo so
as to reproduce the mean and variance of the distribution seen in
data. This scaling is done at the level of the preselection cuts
described in Sections \ref{sec:znn} and \ref{sec:leptonic}. After
the illustrative likelihood cut, the difference in the number of
selected background events and signal efficiencies before and
after re-scaling is used to derive the systematic uncertainty. The
observed variations are then added in quadrature.

\item {\bf Detector tracking resolution}. The tracking parameters for
all Monte Carlo samples are smeared, varying the resolution by
$\pm 10\%$ in $\tan\lambda$ and $\pm 6\%$ in $\kappa$ separately
to evaluate their contributions to the track reconstruction. Here,
$\tan\lambda = \cot\theta$, where $\theta$ is the track polar
angle and $\kappa$ is the track curvature at the point of closest
approach to the origin. The variation sizes are based on a
comparison between data and Monte Carlo using muon pairs and
Bhabha events.

\item {\bf Background cross-section determination}.
The overall 4-fermion cross-section uncertainty is assumed to be
about $\pm 2$\% \cite{bgnd-xs}, reflecting differences in
calculations of the \WW\ and \Zzero\Zzero~cross-sections when
comparing results from various generators. The same error is
assigned to 2-fermion \sm~background. For the electron channel a
20$\%$ error on the cross-section for tagged hadronic two-photon
events is assumed based on a recent OPAL
measurement~\cite{nisius}.

\item{\bf Lepton identification.}
Lepton identification is performed as in~\cite{ww172}. Systematic
errors are assigned to account for observed differences between
data and Monte Carlo simulation in lepton identification and
tracking efficiency. The lepton identification mismodelling is
studied using ``mixed events''. These events are formed by
combining two kinds of events recorded at $\sqrt{s}=91$ GeV: a
$\Zzero \to \qqbar $ event is combined with half of a $\Zzero \to
\ell^+\ell^-$ event to simulate a $\mathrm {W^+W^- \to \mathrm
{qq}\ell\nu}$ event. The systematic error is obtained by comparing
data and Monte Carlo lepton identification efficiency in mixed
events and is estimated to be 0.29\% for electrons and 0.24\% for
muons. The second contribution to the uncertainty accounts for
tracking losses. To determine this contribution, $\Zzero \to \ee$
and $\Zzero \to \mm$ events in data recorded at $\sqrt{s} = 91$
GeV are compared to Monte Carlo events. The difference in the
tracking efficiency is used to extract a 1.1\% systematic error
for electrons and a 0.4\% error for muons. These two contributions
are added in quadrature for each lepton, then doubled since our
selection requires two such leptons per event, leading to a
systematic uncertainty of 2.3\% for electrons and 0.9\% for muons.

\item {\bf Monte Carlo statistics}. The numbers of events
passing the preselection as well as the size of the Monte Carlo
sample before preselection are used to determine the contribution
from statistically limited samples based on binomial statistics.
Table \ref{syst} gives the contributions at $\sqrt{s}= 189$~GeV.
In the limit calculation the systematic errors for each different
final state and centre-of-mass energy are used. \ei

To illustrate the effect of the systematic uncertainties after the
illustrative likelihood cut described in Sections
\ref{sec:likelihood} and \ref{sec:likel}, a signal hypothesis of
\ma~= 6 \gevocc\ and \mh~= 70 \gevocc\ would contribute $38.0 \pm
0.8$ events in addition to the $14.9 \pm 1.5$ events expected from
\sm~backgrounds in the $\Zzero \to \nnbar$ channel after combining
all Monte Carlo samples at 189, 196, 200 and 206 GeV. For the
combined leptonic channels, the signal would contribute $12.9 \pm
0.4$ events in addition to the $7.2 \pm 1.1$ expected events from
backgrounds.
%The expected number of events for one mass
%hypothesis ($\ma = 6$, $\mh = 70$ \gevocc) and the total background is
%given for all data combined
There are 18 candidates selected in the data for the invisible
channel and eight candidates in the leptonic channels as shown in
Tables \ref{cut300}, \ref{cutflowm} and \ref{cutflowe}.
% for the
%$\Zzero \to \nnbar$, \mm\ and \ee\ channels, respectively.

\section{Results}
\label{sec:results} The analyses presented here are designed to
explore the possibilities of a low mass Higgs boson, namely for
\ma~below the \bbbar~threshold. No significant excess of events is
observed in either the invisible or leptonic \Zzero~decay modes.
Hence we set limits within two different scenarios: a
model-independent scenario and the MSSM no-mixing benchmark
parameter scenario. We obtain 95\% confidence level (CL) exclusion
limits using standard statistical procedures based on the
likelihood ratio technique~\cite{scan} as applied in other OPAL
publications. The likelihood variables described in Sections
\ref{sec:likelihood} and \ref{sec:likel} are used as
discriminating variables for the limit calculation without cutting
on this variable. For the signal, the likelihood variable and the
efficiency are calculated for each centre-of-mass energy, each
[\ma,\mh] hypothesis and each final state. The search efficiency
is computed from Monte Carlo samples produced for each \A\ pair
decay channel (namely $\A\A \to$ \ccbar\ccbar, \ttbar\ttbar,
\ggbar\ggbar, \ccbar\ttbar, \ccbar\ggbar, \ttbar\ggbar), each
[\ma, \mh] hypothesis and each centre-of-mass energy considered in
this study. For the \ma\ and \mh\ points located between the mass
points where Monte Carlo samples were generated, the efficiencies,
the shape of the likelihood distribution and the systematic errors
are interpolated using a weighted mean of the relevant quantity
for the four nearest [\ma, \mh] mass points. The likelihood
variable for backgrounds and data are determined separately for
each centre-of-mass energy. The efficiencies are shown in Figure
\ref{eff} at \ecm~=~189 GeV for each \A\A~decay channel versus
\ma\ and \mh, both for the missing energy and the leptonic
analyses. Similar behaviour is observed for any other choice of
the masses and centre-of-mass energy. The efficiencies are
calculated neglecting cases where the \A\ would decay to
resonances. The search is still sensitive to the \A\ decays to
resonances since the resonant states decay preferentially into
\ggbar, \ttbar\ and \ccbar~\cite{reso}.

\subsection{Model-independent limits}
\begin{figure}[H]
\begin{center}
\mbox{ \epsfxsize=\textwidth \epsffile{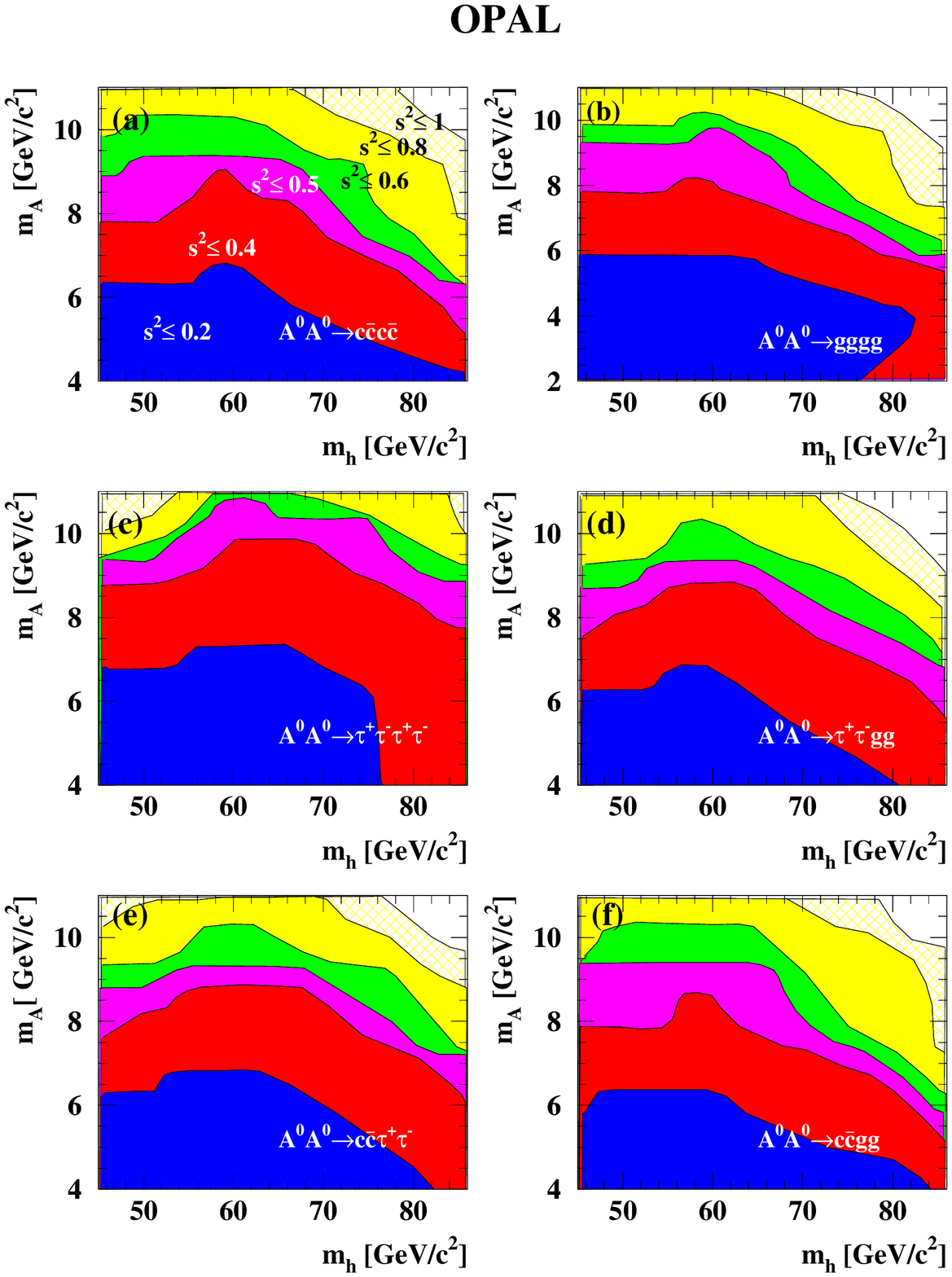}
    }
 \parbox{0.9\textwidth}{\caption {\sl
Upper limits at 95\% CL for $s^2$ in the \ma~versus \mh~plane,
assuming 100\% decays of \h\ into \A\A\ and 100\% decays of \A\A\
into (a) \ccbar\ccbar\, (b) \ggbar\ggbar\, (c) \ttbar\ttbar\, (d)
\ttbar\ggbar\, (e) \ccbar\ttbar\ and (f) \ccbar\ggbar. The
iso-contour lines are for values of $s^2\le$ 1, 0.8, 0.6, 0.5 ,
0.4 and 0.2. These limits are derived using the combined results
from $\Zzero \to \nnbar$, $\Zzero \to \mm$ and $\Zzero \to \ee$
channels and for centre-of-mass energies between 189 and 209 GeV.
\label{modindep}}}
\end{center}
\end{figure}

We calculate limits on the cross-section for the process
$\ee\rightarrow\h\Zzero$. The limits can be extracted in terms of
a scale factor $s^2$ that relates the cross-section for the
production of \h\Zzero, in any specific theoretical interpretation
of our experimental search, to the Standard Model cross-sections:
\beq \sigma_{\mathrm h^0 Z^0} = s^2 \sigma_{\mathrm H^0_{SM} Z^0}.
\eeq The $\h \to \A\A$ branching ratio is assumed to be 100\%. The
limits are extracted for 100\% branching ratio  of \A\A\ into
\ccbar\ccbar, \ggbar\ggbar, \ttbar\ttbar, \ccbar\ggbar,
\ggbar\ttbar\ and \ccbar\ttbar. For each of the six final states
studied, Figure~\ref{modindep} shows the iso-contours of 95\% CL
exclusion for $s^2$ in the \ma~and \mh~mass plane with $2\le \ma
\leq11$ \gevocc\ and 45 \gevocc$\le\mh\le86$~\gevocc. The scan is
performed in 1 \gevocc\ steps in \mh\ and in 0.5 \gevocc\ steps in
\ma. The \ttbar\ttbar\ channel has the largest exclusion power
despite the fact that the selection efficiency is slightly lower
than in the other decay channels since the signal is better
separated from the background.

\subsection{MSSM no-mixing scenario interpretation}
\begin{figure}[htb]
 \begin{center}
   \mbox{
    \epsfxsize=\textwidth
    \epsffile{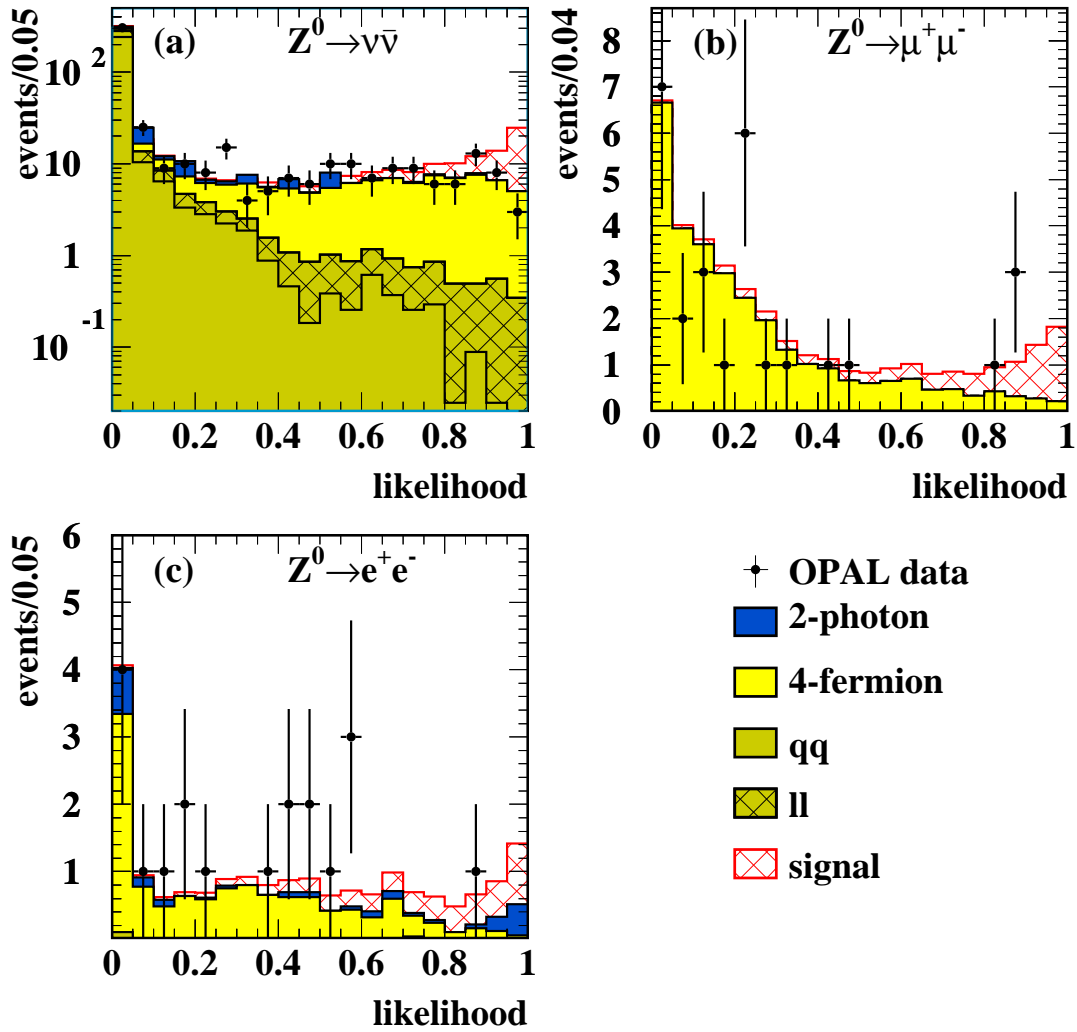}
   }
 \parbox{0.9\textwidth}{\caption {\sl
The likelihood distribution functions in the (a) $\Zzero \to
\nnbar$ (b) $\Zzero \to \mm$ and (c) $\Zzero \to \ee$ channels are
shown for the data, the \sm~backgrounds and the reference signal
(the mixture of all signal hypotheses) for \ecm=189--209 GeV
combined. The backgrounds are added and normalised to the data
integrated luminosity. \label{like300}}}
\end{center}
\end{figure}

\begin{figure}[H]
\begin{center}
\mbox{ \epsfxsize=\textwidth \epsffile{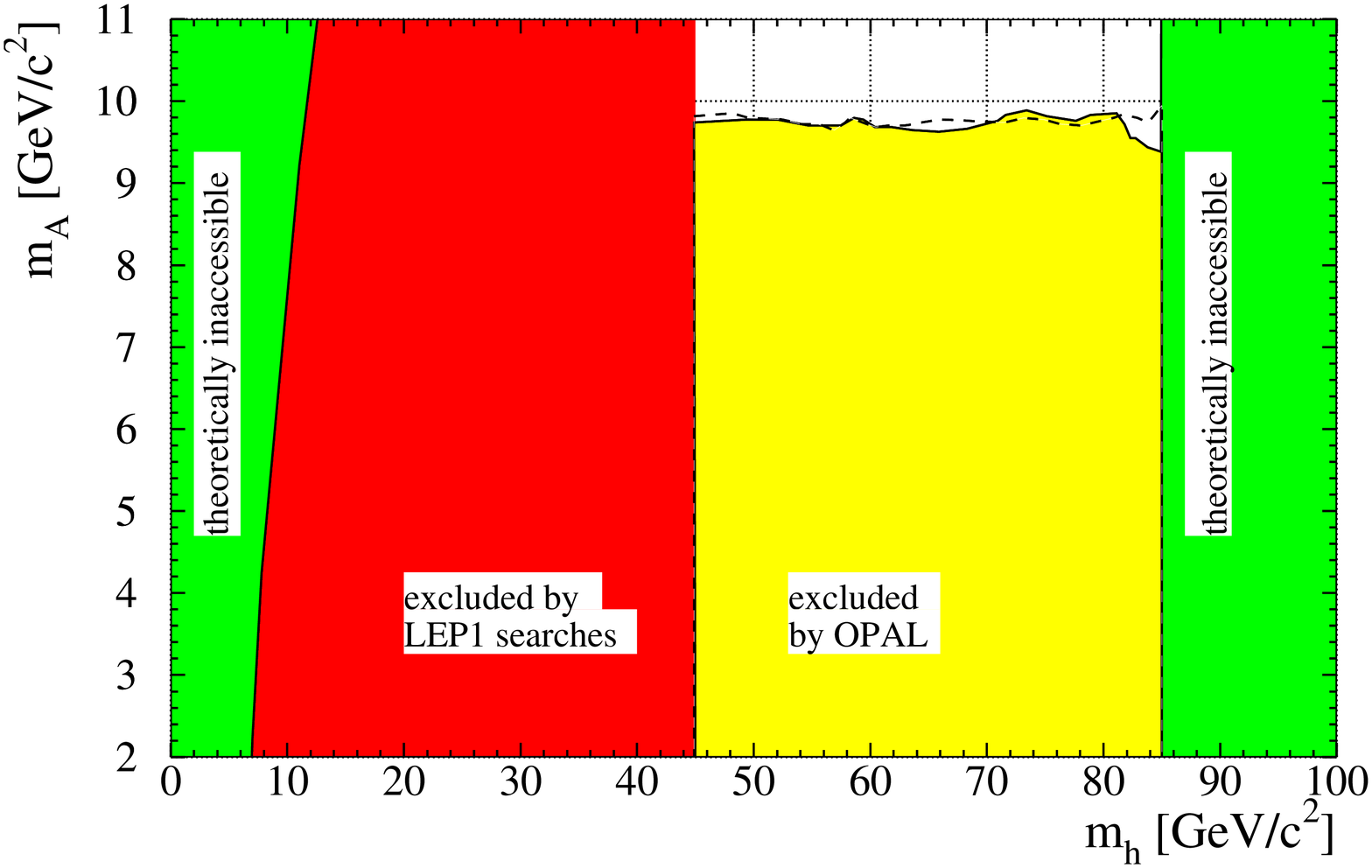}
    }
 \parbox{0.9\textwidth}{\caption {\sl
Expected (dashed contour) and observed (light grey area) excluded
regions at 95\% CL in the \ma~versus \mh~plane for the MSSM
no-mixing benchmark scenario. These limits are derived using the
combined results from $\Zzero \to \nnbar$, $\Zzero \to \mm$ and
$\Zzero \to \ee$ channels and for centre-of-mass energies between
189 and 209 GeV. The theoretically inaccessible regions and the
region excluded by LEP 1 are also shown by darker areas.
\label{exclmssm}}}
\end{center}
\end{figure}

We scan the region with $2\le \ma \le 11$ \gevocc\ and 45
\gevocc$\le\mh\le85$~\gevocc\ in the \ma~versus \mh\ plane for the
MSSM benchmark parameter scenario. The maximum theoretically
allowed value for \mh\ in this scenario is 85
\gevocc~\cite{no-mixing}. The scan procedure is the same as that
of the OPAL MSSM parameter scan~\cite{pr285}. The expected number
of events for the signal is adjusted so as to correspond to
specific production cross-section and branching ratios for a
particular point of the parameter space. The 95\% CL expected and
observed exclusion regions are shown in Figure~\ref{exclmssm}. The
region for $45\le\mh\le82$~\gevocc\ is excluded for $2 \le \ma \le
9.85$ \gevocc, i.e., up to the \bbbar\ threshold where $\A
\rightarrow \bbbar$ decays become dominant. For
$82\le\mh\le85$~\gevocc, the region is excluded for $2 \le \ma \le
9.5$ \gevocc. The whole region below the \bbbar\ threshold was
expected to be excluded but is not due to the presence of
candidates in the missing energy channel (see the third bin from
the right in Figure~\ref{like300} (a)).

\section{Conclusions}
We have searched for the process $\ee\rightarrow\h\Zzero$~with
\Zzero\ decaying into $\nu\bar{\nu},~\ee,~\mm$ and \h\ decaying
into \A\A\ with \ma\ below the \bbbar\ threshold. Six different
decay modes for the \A\A\ system have been investigated:
\ccbar\ccbar, \ggbar\ggbar, \ttbar\ttbar, \ccbar\ggbar,
\ggbar\ttbar and \ccbar\ttbar. No evidence for the presence of a
signal has been found and exclusion limits have been derived both
in a model-independent way and within the MSSM no-mixing benchmark
scenario.

Large areas of the parameter space investigated have been
excluded. In particular, in the MSSM no-mixing scenario, the whole
rectangular area for $2\le\ma\le 9.5$ \gevocc\ and $45\le\mh\le85$
\gevocc\ is excluded at 95\% CL, while the expected exclusion area
is $2\le \ma \le 9.8$ \gevocc\ and $45\le\mh\le 85$ \gevocc. These
limits are the best obtained so far in this region of parameter
space, which has not previously been excluded.

{\bf Acknowledgements:}
\par
We particularly wish to thank the SL Division for the efficient
operation of the LEP accelerator at all energies and for their
close cooperation with our experimental group. In addition to the
support staff at our own institutions we are pleased to acknowledge the  \\
Department of Energy, USA, \\
National Science Foundation, USA, \\
Particle Physics and Astronomy Research Council, UK, \\
Natural Sciences and Engineering Research Council, Canada, \\
Israel Science Foundation, administered by the Israel
Academy of Science and Humanities, \\
Benoziyo Center for High Energy Physics,\\
Japanese Ministry of Education, Culture, Sports, Science and
Technology (MEXT) and a grant under the MEXT International
Science Research Program,\\
Japanese Society for the Promotion of Science (JSPS),\\
German Israeli Bi-national Science Foundation (GIF), \\
Bundesministerium f\"ur Bildung und Forschung, Germany, \\
National Research Council of Canada, \\
Hungarian Foundation for Scientific Research, OTKA T-029328,
and T-038240,\\
Fund for Scientific Research, Flanders, F.W.O.-Vlaanderen, Belgium.\\

%%%%%%%%%%%%%%%%%%%%%%%%%%%

\end{document}